\documentclass[letter]{aa}
\usepackage[dvips]{graphicx}
\usepackage{txfonts}

\def\taucotwo{$\tau_{\rm CO}^{21}$}
\def\tausiotwo{$\tau_{\rm SiO}^{21}$}

\def\tkin{$T_{\rm k}$}
\def\Tco{$T_{\rm ex}^{\rm CO}$}
\def\Tsio{$T_{\rm ex}^{\rm SiO}$}

\def\kms{km~s$^{-1}$}
\def\vlsr{$V_{\rm LSR}$}
\def\Nco{$N_{\rm CO}$}

\def\cmsq{cm$^{-2}$}
\def\cm{cm$^{-3}$}

\begin{document}


\title{High SiO abundance in the HH212 protostellar jet}

\author{S. Cabrit\inst{1}, C. Codella\inst{2,3}, F. Gueth\inst{4}, A. Gusdorf\inst{1}} 

\offprints{S. Cabrit, sylvie.cabrit@obspm.fr}


\institute{
LERMA, UMR 8112 du CNRS, Observatoire de Paris, Ecole Normale Sup\' erieure, Universit\' e Pierre et Marie Curie, 
Universit\' e de Cergy-Pontoise, 61 Av. de l'Observatoire, 75014 Paris, France
\and
INAF, Osservatorio Astrofisico di Arcetri, Largo E. Fermi 5, I-50125 Firenze, Italy
\and
Institut de Plan\' etologie et d'Astrophysique de Grenoble, UMR 5274, UJF-Grenoble 1/ CNRS-INSU, F-38041Grenoble, France
\and
IRAM, 300 Rue de la Piscine, F-38406 Saint Martin d'H\`eres, France}

\date{Received: 08 June 2012; Accepted: 26 October 2012}

\titlerunning{High SiO abundance in the HH212 protostellar jet}
\authorrunning{Cabrit et al.}

\abstract
{Previous SiO maps of the innermost regions of  HH212 set strong constraints on the structure and origin of this jet. They rule out a fast wide-angle wind, and tentatively favor a magneto-centrifugal disk wind launched out to 0.6~AU.}
{We aim to assess the SiO content at the base of the HH212 jet to set an independent constraint on the location of the jet launch zone with respect to the dust sublimation radius.}
{We present the first sub-arcsecond (0\farcs44 $\times$ 0\farcs96) CO map of the HH212 jet base, obtained  with the IRAM Plateau de Bure Interferometer. Combining this with previous SiO(5--4) data, we infer the CO(2--1) opacity and mass-flux in the high-velocity jet and arrive at a much tighter lower limit to the SiO abundance than possible from the (optically thick) SiO emission alone.}
{Gas-phase SiO at high velocity contains at least 10\% of the elemental silicon if the jet is dusty, and at least 40\% if the jet is dust-free, if CO and SiO have similar excitation temperatures. Such a high SiO content is challenging for current chemical models of both dust-free winds and dusty interstellar shocks.}
{Updated chemical models (equatorial dust-free winds, highly magnetized dusty shocks) and observations of higher J CO lines are required to elucidate the dust content and launch radius of the HH212 high-velocity jet.} 


\maketitle

\section{Introduction}
\label{sec:intro}

The jet origin in accreting young stars and its impact on star and planet formation is still a major enigma. Although a magneto-hydrodynamical (MHD) process appears to be required,  it is still debated which fraction of the mass-flux originates from the stellar surface, from reconnections in the magnetosphere-disk interaction zone, or from the disk surface, with all three regions probably contributing to some degree (see eg. Ferreira et al. \cite{ferreira06} for a review). 
Rotation searches in atomic microjets from T Tauri stars set an upper limit of $\simeq$ 0.2--3~AU on the launching radius of any {\it atomic} steady magneto-centrifugal disk wind (Anderson et al. \cite{anderson}, Ray et al. \cite{ray-ppv} and references therein). On the other hand, MHD disk winds launched beyond 0.2--1~AU are expected to be mostly molecular (Panoglou et al. \cite{panoglou}). Therefore the {\it molecular} component of jets holds an important clue to the total radial extent of the launch zone. 

A key species in this respect is SiO,  a specific tracer of molecular jets from the youngest Class 0 protostars, with minimal contamination by ambient swept-up gas (Guilloteau et al. \cite{gui92}, Gueth et al. \cite{gueth99}, Hirano et al. \cite{hirano06,hirano10}). The HH212 jet in Orion (Zinnecker et al. \cite{zinnecker}) provides particularly interesting constraints on jet structure. An inner bipolar microjet peaking at $\pm$ 500--1000~AU from the source was discovered in SiO (Codella et al. \cite{paper1}, hereafter Paper~I). The fastest SiO gas reaches radial velocities $\simeq \pm 10$ \kms\ from systemic, ie. a flow speed $V_p \simeq$ 150 \kms\ after correction for inclination ($\simeq$4\degr\ from the plane of the sky; Claussen et al. \cite{claussen}),  arguing that it is tracing material closely associated with the primary jet. This view is supported by its narrow width $\le$ 90 AU (Cabrit et al. \cite{paper2}, hereafter Paper~II) and by the small blue/red overlap in SiO despite a quasi edge-on view, which restricts the opening angle of the fastest SiO gas to $< $ 4\degr--6\degr\ (Paper~I). At low radial velocities less than $\pm 6$ \kms\ from systemic, SiO traces broader structures consistent with expanding jet-driven bowshocks (Lee et al. \cite{lee-rot}). The pointed shape of these bowshocks requires that the wind speed drops sharply away from the axis. This appears to be inconsistent with the {\it fast} wide-angle wind predicted by the X-wind model (cf. Cai et al. \cite{cai08}),  and requires a {\it slow} wide-angle wind (if any) more compatible with an extended disk wind.

An extended MHD disk wind also seems favored by the tentative rotation signatures reported across the tip of the SiO bowshocks. A magneto-centrifugal launch radius of 0.3--0.05~AU was inferred assuming a flow speed $V_p =$100--200 \kms (Lee et al. \cite{lee-rot}); but the low mean radial velocity of the rotating gas, $|V-V_{\rm sys}|$ = 1.5--4 \kms, implies a lower deprojected flow speed $V_p \simeq 20-60$ \kms, hence a  larger launch radius of 0.6~AU in both lobes (see Equ.~5 in Anderson et al. \cite{anderson}), if the rotation interpretation is correct. Unfortunately, no rotation estimates are yet available for the fastest SiO jet material, as it remains unresolved laterally at 0\farcs3 resolution (Paper~I; Lee et al.  \cite{lee-rot}). 

The SiO gas-phase abundance may offer an independent clue to the location of the jet origin with respect to the dust sublimation radius ($\simeq 0.2$~AU), since silicon is one of the main grain constituents. Our previous line ratio analysis in the HH212 microjet shows that SiO is optically thick for $J_{\rm up}=2$ to 8. The implied minimum SiO abundance strongly depends on the poorly-known jet density and line-of-sight velocity gradient, and varies between 0.05\% and 90\% of the elemental silicon (Paper~II). In this letter, we present new PdBI CO(2--1) observations of the HH212 microjet, which allow us to put more severe constrains on its density, mass-flux, and SiO abundance. We compare the latter with chemical models of dust-free vs. dusty winds, and outline the work required to proceed in solving this question.

\section{Observations}
\label{sec:obs}
\begin{figure*}
\includegraphics[angle=-90,width=9cm]{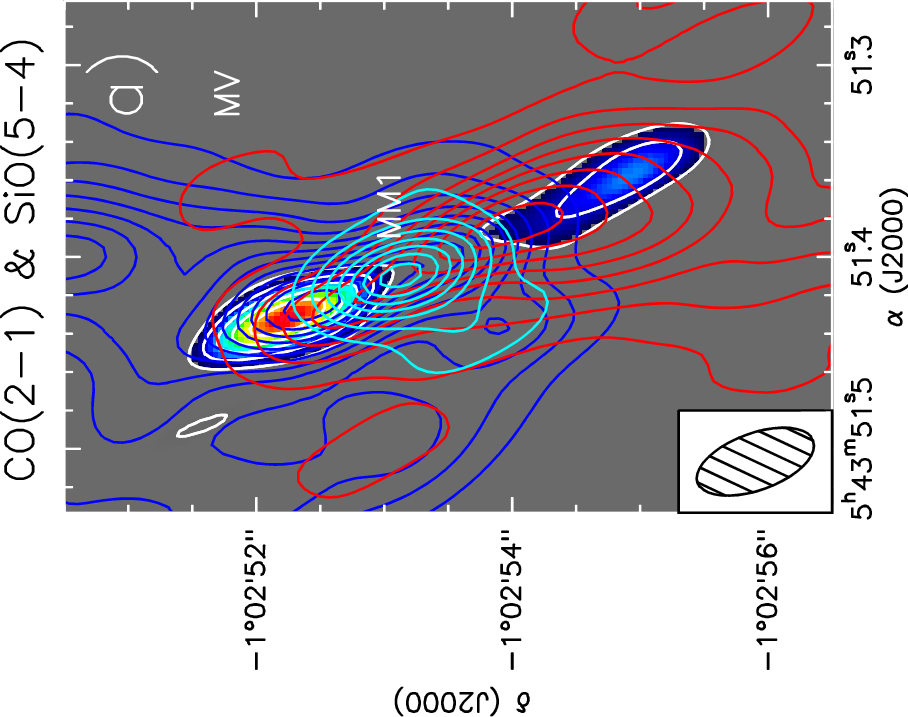}
\includegraphics[angle=-90,width=6.53cm]{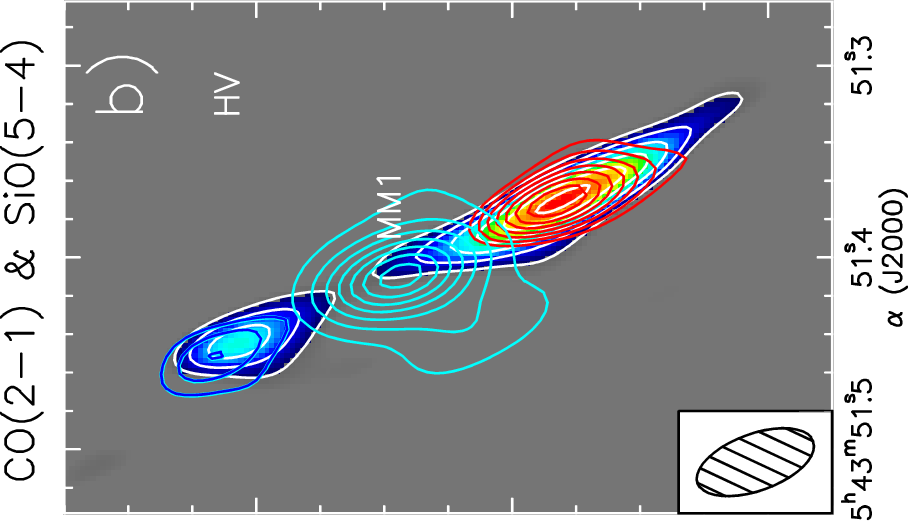}
\caption{CO(2--1) emission (blue and red contours) overlaid on top of  
SiO(5--4) emission from Paper~I (color scale with white contours). 
Both maps were cleaned with the same beam of $0\farcs96\times0\farcs44$ at PA = 20\degr (cf. filled ellipse in lower left corner). 
{\bf a}~{\it (left panel)} Medium-velocity (MV) emission, integrated over [--9.5, --1.5] km s$^{-1}$ (blue), and [+3.5,+9.5] km s$^{-1}$ (red).  
CO contours range from 6$\sigma$ (13~K~km s$^{-1}$) to 78$\sigma$ by 6$\sigma$.
SiO contours range from 6$\sigma$ (74~K~km s$^{-1}$) to 18$\sigma$ by 3$\sigma$.
{\bf b}~{\it (right panel)} High-velocity (HV) emission integrated over [--16.5,--9.5] km s$^{-1}$ (blue), and [+9.5,+12.5] km s$^{-1}$ (red).
CO contours range from 6$\sigma$ (1.4  and 1.1~K~km s$^{-1}$ for blue and red, respectively)
to 13$\sigma$ by 1$\sigma$. SiO contours range from 6$\sigma$ (21.5~K~km s$^{-1}$)
to 21$\sigma$ by 3$\sigma$. 
In both panels, 1.4mm continuum from the central source MM1 is shown in light-blue contours, ranging from 10$\sigma$ (5 mJy beam$^{-1}$) to 70$\sigma$ by steps of 10$\sigma$.
}
\label{fig:maps}
\end{figure*}

CO(2--1) observations of the central region of the HH212 outflow were obtained 
in February 2008 and January, March, and April 2009 
with the IRAM Plateau de Bure Interferometer (PdBI) in France.
The six-element array was used in its A, B, and C configurations 
(baselines from 48~m up to 760~m) for a total time of $\sim$23 hours.
The CO(2--1) line at 230.5380 GHz was observed with a 40 MHz ($\sim$ 50 km s$^{-1}$) bandwidth and a channel sampling of $\sim$ 0.1 km s$^{-1}$, later smoothed to 1 km s$^{-1}$ to increase the signal-to-noise ratio. Two units with a bandwidth of 320 MHz were used to measure the continuum. 
The data were reduced using the public GILDAS\footnote{http://www.iram.fr/IRAMFR/GILDAS} 
software. Amplitude and phase were calibrated on observations of 0528+134 and 0605--085. 
The absolute flux density scale was determined on MWC\,349 
with an estimated uncertainty of $\sim$ 25\%.  
Line and continuum maps were produced using natural weighting and were restored with a clean beam of $0\farcs96\times0\farcs44$ (PA=20$^\circ$). Since we are interested only in the innermost jet knots within $\pm$2\arcsec\ of the central source, no correction was applied for primary beam attenuation (HPBW $\simeq$ 22\arcsec). 
The flux, position, and size of the continuum source all agree with the 1.4mm  data 
presented in Paper~I, except for the tentative source MM2, which is not confirmed. The secondary source seen at 850$\mu$m by Lee et al. (\cite{lee-rot}) at $\Delta\alpha$ = +1\arcsec, $\Delta\delta$ = --0\farcs7 is also unconfirmed (see Fig.~\ref{fig:maps}). In the following, we concentrate on our results in CO(2--1). Radial velocities are expressed in the $V_{\rm LSR}$ rest frame unless otherwise specified. 

\section{Results}

\subsection{The inner 5\arcsec: outflow cavity vs. jet emission}

Figure~\ref{fig:maps} compares in two different velocity ranges the cleaned CO(2--1) map of the inner
5\arcsec\ of the outflow with the SiO(5--4) map from Paper~I, restored here with the same synthesized beam as for CO(2--1). 
In the {\it left-hand panel} (Fig.~\ref{fig:maps}a) both lines are summed over a medium-velocity interval (hereafter MV) covering [$-9.5$\kms,$-1.5$\kms] in the blue and [$+4.5$\kms,$+9.5$\kms] in the red. It can be seen that the CO MV emission peaks along the SiO axis, but also delineates a broader component not seen in SiO, with a biconical morphology opening away from the exciting source. This corresponds to the base of the CO swept-up cavity mapped on larger scale by Lee et al. (\cite{lee-13co, lee-co32}). Owing to this extended component and  the lack of short spacings in our $u-v$ coverage, the reconstructed CO brightness in the MV range is very sensitive to the chosen cleaning parameters.
In the {\it right-hand panel} (Fig.~\ref{fig:maps}b) both lines are summed over a high-velocity interval (hereafter HV) covering [$-16.5$\kms,$-9.5$\kms] in the blue and [$+9.5$\kms,$+12.5$\kms] in the red. In this HV interval, CO is seen to trace the same gas as SiO, ie. a narrow jet still unresolved in the transverse direction, with no significant contamination from the broad cavity.  We verified that contrary to the MV emission, the HV emission is compact enough that its flux is robust and independent of cleaning parameters. 

\subsection{CO(2--1) opacity in the HV jet}
\label{sec:tauco}

As shown in Paper~II and confirmed by Lee et al. (\cite{lee-rot}), SiO(5--4) is close to LTE and optically thick in the inner HH212 knots. On the other hand,  the high ratio CO(3--2)/CO(2--1) $> 1.5-2$ seen at high velocity (Lee et al. \cite{lee-co32}) requires that CO(2--1) is optically thin in the HV jet. We may then infer its opacity \taucotwo\ from the line temperature ratio $R$ of CO(2--1) to SiO(5--4) as $R \simeq \tau_{\rm CO}^{21}$(\Tco/\Tsio), where \Tco\ and \Tsio\ are the excitation temperatures of CO and SiO, respectively. Since CO has a lower dipole moment than SiO, it is even more easily at LTE and we expect that \Tco\ $\simeq$ \tkin. Furthermore, the SiO knots are sufficiently far from the protostar (500 AU) and from any other radiation source that mechanical heating (eg. shocks, mixing-layers, ambipolar diffusion) will largely dominate over radiative heating (Shang et al. \cite{shang-xrays}, Panoglou et al. \cite{panoglou}). Therefore, any strong local gradient in \tkin\ should come with a steep accompanying velocity gradient, and we will minimize the difference between \Tco\ and \Tsio\ caused by such gradients by examining the ratio $R$ {\it as a function of velocity}.  

\begin{figure}
\includegraphics[angle=-90,width=7cm]{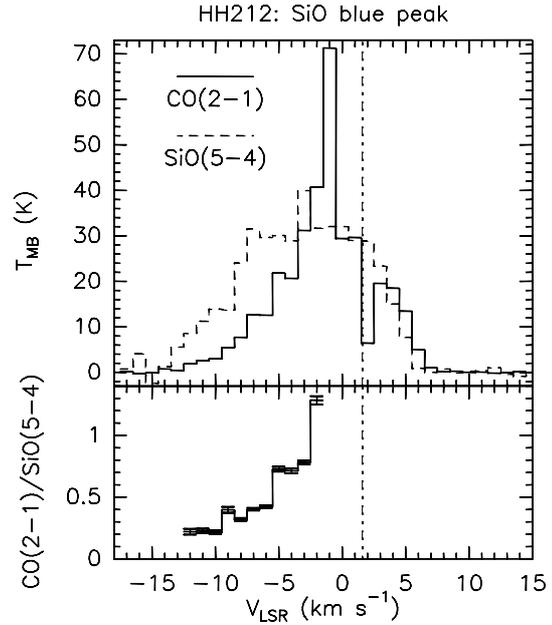}
\caption{{\it Top panel:} Line profiles in CO(2--1) (solid) and SiO(5--4) (dashed) toward the northern SiO peak. {\it Bottom panel:} Ratio of CO(2-1) to SiO(5--4). The vertical 
dot-dashed line marks the ambient LSR velocity (+1.6 km s$^{-1}$; Wiseman et al. \cite{wiseman}). 
}
\label{fig:profile}
\end{figure}

In Figure~\ref{fig:profile}, we plot the main beam temperature ratio $R$ as a function of velocity toward the (northern) blue SiO inner knot. In the HV blue range $V < -9.5$\kms, the CO(2--1)/SiO(5--4) ratio is characterized by a low value $R \simeq$ 0.2, and we infer that \taucotwo (\Tco/\Tsio) $\simeq 0.2$. In the redshifted HV range, the ratio $R$ appears similar but CO(2--1) is affected by a strong dip around \vlsr $\simeq +9$ \kms, also seen in CO(3-2) (Lee et al. \cite{lee-co32}). This dip is present at all positions and may be caused by an extended foreground component fully resolved-out by interferometers.
Given this complicating circumstance, we focus on the blue HV jet below. 

\subsection{HV jet density and mass-flux}
\label{sec:mdot}

Since the HV blue material is optically thin (Lee et al. \cite{lee-co32}), its beam-averaged CO column density, \Nco(HVB), can be inferred from the CO(2--1) integrated intensity, without  opacity correction. At the northern peak, we measure $T_{\rm mb}dV = 8.2$ K \kms\ over the HV range and infer that  \Nco(HVB) $\simeq 5 \times 10^{15} (T_{\rm ex}^{\rm CO} / 50K)$ \cmsq\  for LTE in the range $\simeq 50-500$~K  determined in Paper~II. 

We may also infer a lower limit to the volume density of the HV blue jet at the northern peak. With a jet diameter $D < 0\farcs2 = 90$AU (cf. Paper~II), and a beam size across the jet $b_\perp = 0\farcs44 = 200$~AU, we find $n_{\rm H} \simeq  (b_\perp / D^2) N_{\rm CO} / X_{\rm CO} > 10^5 (T_{\rm ex}^{\rm CO} / 50K) \times (10^{-4}/ X_{\rm CO}) $\cm, where $X_{\rm CO}$ is the CO gas-phase abundance relative to H nuclei. 

Assuming a steady flow along the jet axis, we may also estimate the {\it one-sided} jet mass-flux as given by  (see Lee et al. \cite{lee-co32})
$\dot{M}_j = 1.4 m_H  V_j b_\perp N_{\rm CO} X_{\rm CO}^{-1}$ 
$\simeq$ 
$ 10^{-7} \left( {T_{\rm ex}^{\rm CO} / 50 {\rm K}} \right) \left( { 10^{-4} / X_{\rm CO}}  \right) M_\odot {\rm yr}^{-1}, $
where  $V_j = 150$ \kms\ is the relevant deprojected speed for the HV jet (cf. Sect.~\ref{sec:intro}). 
This $\dot{M}_j $ is ten times lower than the value of Lee et al. (\cite{lee-co32}) for the same \Tco, because we 
excluded the MV CO emission (dominated by the slower bowshocks and the large-scale cavity). 
The appropriate choice of  $X_{\rm CO}$ depends on the assumed dust content in the jet. Chemical models of dense Class 0 jets show that CO self-screens efficiently against FUV photons and contains most of the available gas-phase carbon not locked in grains (Glassgold et al. \cite{glassgold}, Panoglou et al. \cite{panoglou}). Correspondingly, we expect $X_{\rm CO} \simeq 3.5 \times 10^{-4}$ (solar carbon abundance; Holweger \cite{solar}) when the wind is dust-free, and  $X_{\rm CO}$ $\simeq 10^{-4}$ for an interstellar dust/gas ratio $\simeq$ 1\%  (Flower \& Pineau des For\^ets \cite{FPdF03}). Hence, the above values for $\dot{M}_j$ and $n_H$ are decreased by a factor $\simeq 3.5$ if the HV jet is grain-free, while they both increase by a factor 10 for a high \Tco\ = 500~K. 

\subsection{SiO abundance in the high-velocity jet}
\label{sec:xsio}

Using the on line RADEX code (Van der Tak et al. \cite{radex})  in the large velocity gradient (LVG) approximation, we  compared the LTE opacities of SiO lines versus that of CO(2--1). 
For the SiO(2--1) line and  \tkin\ = 50--500~K, we find that
\begin{equation}
{X_{\rm SiO} \over X_{\rm CO}} = {N({\rm SiO})/\Delta V \over N({\rm CO})/\Delta V} \simeq {{\tau_{\rm SiO}^{21} /  \tau_{\rm CO}^{21}} \over (122\pm 10) }  \left({T_{\rm ex}^{\rm SiO} \over  T_{\rm ex}^{\rm CO}}\right)^{1.9}. 
\label{eq:xsio}
\end{equation}
With our result from Sect.~\ref{sec:tauco} that \taucotwo  (\Tco/\Tsio) $\simeq R = 0.2$, and the observational constraint \tausiotwo $> 1$ (cf. Paper~II), we infer that $X_{\rm SiO}$/$X_{\rm CO} > 0.04$f (\Tsio/\Tco)$^{0.9}$.
Adopting the $X_{\rm CO}$ values discussed above and a solar elemental abundance for silicon [Si/H]$_\odot \simeq 3.5\times 10^{-5}$ (Holweger,  \cite{solar}), we obtain that gas-phase SiO in the inner  blue jet knot of HH~212 represents  {\it at least  $\simeq$ 10\%} of the elemental silicon if the wind is dusty, and {\it at least $\simeq$ 40\%} of it if the wind is dust-free,  with an uncertainty factor of (\Tsio/\Tco)$^{0.9}$. The uncertainty in $X_{\rm SiO}$ from this method is $\la$ factor 2 in planar C-shock models. 

\section{Implications for the launch radius of the SiO jet}

An SiO abundance reaching $\ge$ 10\%-40\% of elemental silicon is challenging for current models of MHD winds --- both dust-free and dusty. We discuss each of these two cases below. 

The chemistry of {\it dust-free} MHD winds was investigated in 1D by Glassgold et al. (\cite{glassgold}).
The wind is launched from a 5000~K protostar of radius $R_0 = 10 R_\odot$ = 0.045~AU, and undergoes various degrees of acceleration and expansion. The fraction of silicon in the form of SiO reaches  $\ga$ 50\% only when the base density exceeds a critical value, corresponding to an isotropic mass-flux rate $\ge 3 \times 10^{-6} M_\odot$/yr in their accelerating model. This is well above our estimate of $\dot{M}_j \le 3\times 10^{-7}  M_\odot {\rm yr}^{-1}$ for the HV jet in HH212, if dust free with \tkin $\le$ 500~K (see Sect.~\ref{sec:mdot}).
Furthermore, when a flat far-ultraviolet (FUV) excess below 2000\AA\ is present, SiO is entirely photodissociated throughout the wind (see Fig.~10 of Glassgold et al. \cite{glassgold}). The assumed  FUV photon flux at the SiO dissociation limit $\simeq$8eV is similar to that produced by an accretion shock of blackbody temperature 10,000~K and total luminosity 3.5$L_\odot$. The Class 0 protostar of 14$L_\odot$ driving the HH212 outflow (Claussen et al. \cite{claussen}) should have a FUV flux from accretion at least as high, hence negligible SiO in the wind. 
Increasing the launch radius would not solve the problem, since both the base wind density and the radiation field would drop as $1/R_0^2$, keeping the same ratio of SiO reformation to photodissociation rates. Therefore, spherical dust-free winds do not seem able to reproduce the high SiO content in the HH212 jet.

A possible factor favoring SiO synthesis in dust-free winds 
would be if the jet does not arise from $4\pi R_0^2$ but from a narrow equatorial annulus of width $\Delta R \ll R_0$, eg. as assumed in the X-wind model (Cai et al. \cite{cai08}). This would increase the base density by a large factor $R_0/\Delta R$ for the same jet mass-flux, thus enhancing Si$^+$ recombination and SiO formation with respect to photodissociation. However, this hollow wind geometry would also enable the penetration of destructive FUV photons through the emptied polar regions (see Panoglou et al. \cite{panoglou}), an effect not included in the 1D calculations of Glassgold et al. (\cite{glassgold}). 
Detailed chemical wind models in 2D with a FUV accretion excess are thus necessary to check whether an equatorial dust-free wind could reach an SiO abundance  $\ga$ 40\% of elemental Si, as estimated in the HV jet of HH~212. 


An alternative scenario is that SiO would form by shock processing of silicate grains in a {\it dusty wind} launched from the disk (Chandler et al. \cite{chandler}; Panoglou et al. \cite{panoglou}). Recent models of C-type shocks with ion-neutral decoupling suggest that sputtering of charged grains by drifting neutrals can release up to 5\% of Si in the gas phase, which then reacts with O$_2$ or OH to form SiO (Schilke et al. \cite{schilke}, Gusdorf et al. \cite{gusdorf08a}). 
A possible problem with sputtering, raised in Paper~II, is that the dynamical timescale 
for the inner SiO knots in HH212 is only 25 yr, which may be too short to complete the sputtering process and the conversion of Si into SiO (Gusdorf et al. \cite{gusdorf08b}). Fortunately, at the high densities $\ge 10^5$\cm\ of the HH212 inner knots (cf. Sect.~\ref{sec:mdot}), 
grain-grain collisions also become important (Caselli et al. \cite{caselli}). They will alleviate this problem in two ways:  First, by shattering dust grains into smaller fragments, they increase the coupling between charged and neutral fluids and shorten the C-shock timescale. Second, they induce grain vaporization, which directly releases SiO molecules into the gas-phase, reaching up to 5\% of the total silicon at 40 \kms\ (see Guillet et al. \cite{guillet-cshock}). The same process may also occur in single-fluid (J-type) shocks, with SiO reaching 2\% of the total silicon at 50 \kms\ (Guillet et al. \cite{guillet-jshock}). We note that the above SiO yields refer only to  {\it interstellar} shock models with Alfv\' en speeds $V_A $ of a
few \kms, a value $\simeq 10$ times lower than expected at 500~AU in protostellar jets (Hartigan et al. \cite{harti-bfield}, Garcia et al. \cite{garcia1a}). Extrapolating the trends with $V_A$ shown in Fig.~5 of Caselli et al. (\cite{caselli}) and Fig.~6 of  Guillet et al. (\cite{guillet-jshock}) suggests that in dusty MHD jets, grain-grain collisions might be able to release $\ga$ 10\% of Si in the form of SiO, compatible with our estimate in HH212 in the dusty wind case. Dedicated shock models with appropriate initial conditions would be necessary to verify this extrapolation. 

In conclusion,  $\ga$ 10\%--40\% of elemental silicon in the form of SiO is challenging for current chemical models of both dust-free winds and dusty shocks, and calls for additional modeling with more realistic assumptions to clarify the maximum SiO abundance that can be reached in either case. ALMA observations in CO(3--2) and CO(6--5) will also be essential for better constraining  the jet temperature, width,  mass-flux, and density, as well as the ratio \Tsio/\Tco\ entering our  SiO abundance estimate, all of which are key parameters for distinguishing among current MHD jet  models.



\begin{acknowledgements}
We are grateful to G. Pineau des For\^ets and M. Tafalla for useful discussions, to an anonymous referee for comments that helped to strengthen the paper, and to the IRAM staff for their support with observations. This research has made use of NASA's Astrophysics Data System, and received financial support from  the Programme National de Physico-Chimie du Milieu Interstellaire (PCMI). 
The IRAM Plateau de Bure Interferometer is funded by INSU/CNRS (France), MPG (Germany) and IGN (Spain).
\end{acknowledgements}






\end{document}